\documentclass[letter]{aa} 

\usepackage{graphicx}
\usepackage{txfonts}
\usepackage{multirow}
\usepackage{hyperref}
\usepackage{color}
\newcommand{\orcid}[1]{\href{https://orcid.org/#1}{\includegraphics[width=10pt]{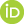}}}

\begin{document} 

    \titlerunning{Giant outbursts of clumpy material preceding Type II SN 2024qiw}
    \authorrunning{T. Nagao et al.}

   \title{Giant outbursts of clumpy material preceding Type II supernova 2024qiw
   }

   \author{T.~Nagao\inst{1,2,3,4 \orcid{0000-0002-3933-7861}}\fnmsep\thanks{takashi.nagao@nao.ac.jp}
          \and
          H.~Kuncarayakti \inst{1,5\orcid{0000-0002-1132-1366}}
          \and
          K.~Maeda \inst{6\orcid{0000-0003-2611-7269}}
          \and
          S.~Mattila, \inst{1,7\orcid{0000-0001-7497-2994}}
          \and
          R.~Kotak, \inst{1}
          \and
          T.~Killestein \inst{8\orcid{0000-0002-0440-9597}}
          \and
          C.~Humina, \inst{1}
          \and
          D.~Steeghs \inst{8\orcid{0000-0003-0771-4746}}
          \and
          D.~Jarvis \inst{9}
          }

   \institute{Department of Physics and Astronomy, University of Turku, FI-20014 Turku, Finland\\
   \email{takashi.nagao@utu.fi}
         \and
         Aalto University Mets\"ahovi Radio Observatory, Mets\"ahovintie 114, 02540 Kylm\"al\"a, Finland
         \and
         Aalto University Department of Electronics and Nanoengineering, P.O. BOX 15500, FI-00076 AALTO, Finland
         \and
         National Astronomical Observatory of Japan, National Institutes of Natural Sciences, 2-21-1 Osawa, Mitaka, Tokyo 181-8588, Japan
         \and
         Finnish Centre for Astronomy with ESO (FINCA), FI-20014 University of Turku, Finland
         \and
         Department of Astronomy, Kyoto University, Kitashirakawa-Oiwake-cho, Sakyo-ku, Kyoto 606-8502, Japan
         \and
         School of Sciences, European University Cyprus, Diogenes Street, Engomi, 1516, Nicosia, Cyprus
         \and
         Department of Physics, University of Warwick, Gibbet Hill Road, Coventry CV4 7AL, UK
         \and
         Astrophysics Research Cluster, School of Mathematical and Physical Sciences,
University of Sheffield, Sheffield S3 7RH, UK
             }

   \date{Received ?? ??, ??; accepted ?? ??, ??}

 
  \abstract{
  Observations of core-collapse supernovae suggest that some massive stars undergo intense mass loss shortly before explosion, but the underlying mechanisms remain unknown. Here we report evidence of giant outbursts of clumpy material from a massive star in the final decades before explosion. Photometric, spectroscopic, and polarimetric data of SN~2024qiw reveal a bumpy light curve, a broad H$\alpha$ profile, and variable polarization, all consistent with interaction between SN ejecta and clumpy circumstellar material, implying a mass-loss rate of $\gtrsim 10^{-2}$ M$_\odot$ yr$^{-1}$. Taken together, the most likely explanation is multiple major eruptions, similar to those of Luminous Blue Variables (LBVs), but occurring shortly before explosion. This challenges standard stellar evolution theory by requiring either that LBVs explode terminally, or that other evolutionary phases produce eruptive episodes. In spite of very high pre-SN mass loss, the resulting SN is of Type~II, rather than Type IIn, highlighting diverse and previously unrecognized late-stage mass-loss processes.
  }


   \maketitle
%

\section{Introduction} \label{sec:introduction}

Massive stars ($\gtrsim 8$ M$_{\odot}$) end their lives as core-collapse supernovae (SNe), yet their final evolutionary stages remain one of the unsolved problems in the well-established stellar evolution theory \citep[e.g.,][]{Kippenhahn2013}. Recent supernova (SN) observations have suggested previously unknown evolutionary processes accompanied by extensive mass ejections during the final phases of massive stars. The majority of Type II SNe show evidence for the dense circumstellar material (CSM; the so-called confined CSM, corresponding to a mass-loss rate of $\gtrsim 10^{-3}$ M$_{\odot}$ yr$^{-1}$), as indicated by flash ionization lines in the spectra \citep[e.g.,][]{Khazov2016, Yaron2017, Bruch2021}, rapid rises in their light curves \citep[LCs; e.g.,][]{Forster2018}, and/or high polarization at early phases \citep[][]{Vasylyev2023}. In some extreme Type IIn SNe, which show narrow hydrogen emission lines from slow-moving CSM in their spectra, even several tens of solar masses of CSM are required to explain their luminous, long-lived radiation \citep[e.g.,][for reviews]{Smith2017a, Fraser2020, Dessart2024}. 
The mechanisms behind these mass ejections remain unclear. The inferred CSM properties imply that these mass ejections far exceed the well-understood stellar winds, e.g., the line-driven wind. Nevertheless, several potential ideas have been proposed to explain such extensive mass loss, including mass loss due to binary interaction \citep[e.g.][]{Podsiadlowski1992, Chevalier2012, Ouchi2017}, mass loss driven by energy input from wave excitation caused by core convection \citep[e.g.,][]{Quataert2012, Fuller2017}, mass loss resulting from internal stellar instabilities \citep[e.g.,][]{Arnett2011, Woosley2015} and mass loss due to eruptions of unstable stars such as Luminous Blue Variables (LBVs) \citep[e.g.,][]{Humphreys1994, Kotak2006, Harpaz2009, Smith2017b, Smith2026}.

In this paper, we present our photometric, spectroscopic, and polarimetric observations of the peculiar Type II SN~2024qiw. Details of the SN — including the discovery date, estimated explosion epoch, redshift, and adopted extinction — are provided in Appendix~\ref{sec:appA}, while the observational data and their reduction processes are described in Appendix~\ref{sec:appB}. 

\section{Observational properties} \label{sec:result}

\begin{figure}
   \centering
            \includegraphics[width=\hsize]{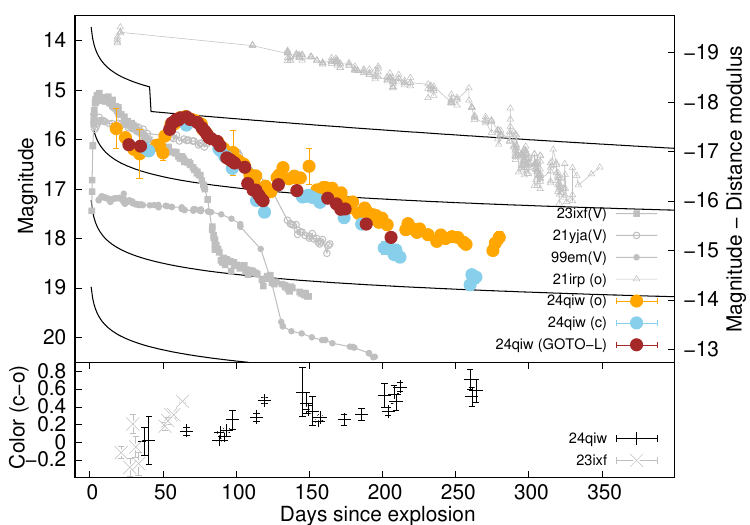}
      \caption{Photometry of SN~2024qiw. Top: optical LCs (colored points) compared with those of other SNe (gray points; plotted toward the right $y$-axis). The comparison data are from \citet[][]{Singh2024,Anderson2014, Galbany2016, Nagao2024, Reynolds2025a}. The black lines show $V$-band LCs from the CSM-interaction model of \citet[][]{Moriya2013}, assuming SN ejecta with $n=12$, $\delta=1$, $M_{\rm{ej}}=10$ $M_{\odot}$, and $E_{\rm{ej}}=10^{51}$ erg, and a spherical CSM with $v_{\rm{w}}=100$ km s$^{-1}$ and $s=2$, and a convergent efficiency $\epsilon=0.5$. The lines correspond to mass-loss rates of $10^{-1}$ to $10^{-4}$ $M_{\odot}$ yr$^{-1}$ (top to bottom). Bottom: the $c$-$o$ evolution compared with SN~2023ixf.
              }
      \label{fig:LC}
   \end{figure}

Fig.~\ref{fig:LC} shows the optical light curves of SN~2024qiw, compared with those of several Type II SNe, including a Type IIL SN (2023ixf), a bright Type IIP SN (2021yja), a faint Type IIP SN (1999em), and a bright long-lived Type II SN (2021irp). SN~2024qiw shows unique long-lasting light curves with bumpy structures. It does not have a clear drop from the plateau to the tail phase as in Type II SNe. Although the actual peak magnitude of SN~2024qiw is not very certain due to the lack of observations before the first detection, the first detection ($o\sim -17.5$ mag) should be around the peak brightness based on the spectral comparison with another Type~II SN, i.e., SN~2023ixf (see Appendix~\ref{sec:appA}). During the first hundred days, SN~2024qiw is brighter than the relatively faint SN 1999em and comparable to the brighter SN~2021yja. At late phases from $\sim 100$ days after the explosion, it is much brighter than SN~1999em and slightly brighter than SN~2021yja. During the first $\sim35$ days, SN~2024qiw shows a relatively fast decline in the $o$-band light curve, as seen in Type~IIL SNe rather than in Type~IIP SNe. In particular, its photometric and spectroscopic evolutions at these early phases are very similar to those of the Type IIL SN 2023ixf (see also Appendix~\ref{sec:appC}). The light curves of SN~2024qiw show multiple clear peaks at phases $\lesssim 20$, $\sim65$ and $\sim130$ days after the explosion. The degree of the brightening is different during the first and the second rebrightening episodes. During the first rebrightening (around $\sim 50$ days), $c-o$ is around 0.1 mag. Then, from $\sim 100$ days, the color becomes redder with $c-o \sim 0.4$ mag. Finally, the $c-o$ color becomes $\sim 0.7$ mag at $\sim 200$ days.

\begin{figure}
   \centering
            \includegraphics[width=\hsize]{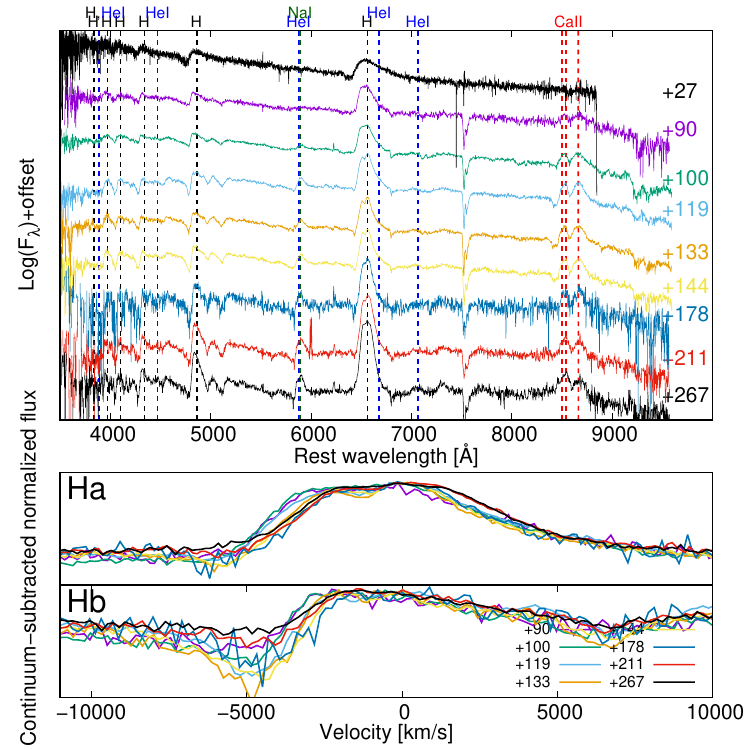}
      \caption{Spectral sequence of SN~2024qiw (top panel) and evolution of the H$\alpha$ and H$\beta$ line profiles (bottom panel). The continuum level is determined by the fluxes at the wavelength ranges from -20000 to -15000 km s$^{-1}$ and from 9000 to 11000 km s$^{-1}$. The phases are given relative to the explosion.
              }
      \label{fig:spec}
   \end{figure}

Fig.~\ref{fig:spec} shows the spectral evolution of SN~2024qiw. The photospheric spectra show a continuum and emission lines from allowed lines (e.g., Balmer lines, He I $\lambda$5876, Na I D $\lambda\lambda$ 5890, 5896, and/or Ca II NIR triplet $\lambda\lambda\lambda$ 8498, 8542, 8662) for at least $\sim 250$ days. At later epochs, the so-called Fe bump (i.e., excessive fluxes at $\lambda \lesssim 5500$ {\AA}, which is probably due to ionized/excited Fe lines) emerges. The general features such as the continuum and broad Balmer lines are similar to the spectra of Type~IIP/L SNe, while the long photospheric phase and the emission-dominated line shapes of Balmer lines resemble those of 21irp-like SNe, which are Type II SNe powered by disk-CSM interaction \citep[e.g.,][see also Appendix~\ref{sec:appC}]{Reynolds2025a,Nagao2025}. 

The evolution of the H$\alpha$ line shape is correlated with the light-curve evolution. As time progresses from the first to the second rebrightening, a component around $\sim +2000$ km s$^{-1}$ gradually emerges, with a reduction of the component around $\sim -4000$ km s$^{-1}$. It is difficult to see this behavior in the H$\beta$ profile, likely due to the stronger absorption and the lower signal-to-noise ratio. In addition, the strength of the H$\alpha$ absorption becomes peaked around the beginning of the second rebrightening, which is also seen in the H$\beta$ line. This is a clear sign that the emitting regions became deeper inside the SN ejecta toward the observer during this transition.

\section{Discussion}\label{sec:discussion}

\begin{figure}
   \centering
            \includegraphics[width=\hsize]{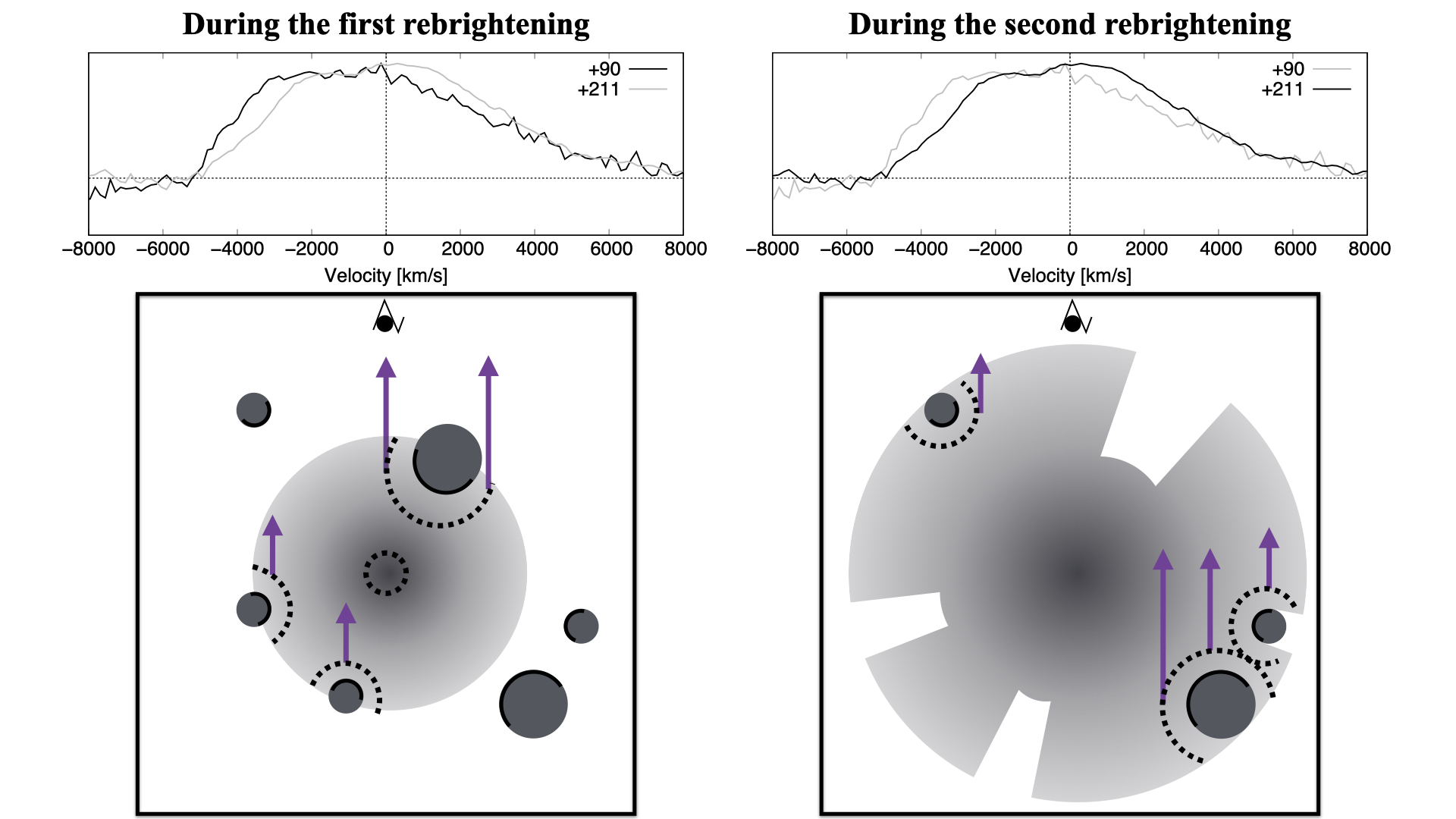}
      \caption{Schematic illustration of our interpretation of the observational properties of SN~2024qiw. The left and right panels correspond to the first and second rebrightenings, respectively, with the system viewed from above each bottom panel. The enhancement of the $-4000$km/s component in the H$\alpha$ profile during the first rebrightening indicates interaction with a major CSM clump on the near side of the SN ejecta, while the enhancement of the $+2000$ km/s component during the second rebrightening suggests interaction with another major clump on the far side. The dotted lines show the locations of the local photospheres produced by the CSM-clump interaction.
              }
      \label{fig:schematic_picture}
   \end{figure}

The bumpy light curve of SN~2024qiw suggests some episodic energy injections in the SN ejecta at least twice, in addition to the actual core-collapse explosion. This fact already disfavors the possibility of a large amount of $^{56}$Ni or a central engine \citep[e.g., magnetar, fallback accretion;][]{Kasen2010, Dexter2013} as the explanation for these rebrightenings, as such energy sources cannot create multiple major episodic energy injections with a time interval of $\sim 100$ days. In addition, the spectroscopic properties of SN~2024qiw do not support these scenarios \citep[e.g.,][]{Dessart2018}. A more reasonable explanation for the bumpy LC would be interaction of the SN ejecta with clumpy or multiple-shell CSM. The CSM-interaction scenario is further supported by the spectral similarity to other SNe interacting with aspherically distributed CSM, such as SN~2021irp, which was powered by the interaction between the SN ejecta and a massive, disk-like CSM \citep[][]{Reynolds2025b}. In particular, the slowly evolving spectra with broad Balmer lines are similar to those shown by SN~2021irp (see Appendix~\ref{sec:appC}). This can be interpreted as evidence for the main radiation originating from local photospheres in the outer layers of the SN ejecta, formed by heating from local interaction regions hidden by the ejecta (see Fig.~\ref{fig:schematic_picture}), as suggested in the case of SN~2021irp \citep[see also the discussions in][]{Reynolds2025b}.

The observations suggest a clumpy, rather than multiple-shell, CSM distribution. This is supported by clear correlations among the light-curve evolution, the H$\alpha$ line profile, and the color changes. During the transition from the first to second rebrightening, the component around $\sim +2000$ km s$^{-1}$ in the H$\alpha$ line increases, while the one at $\sim -4000$ km s$^{-1}$ weakens, and the color becomes progressively redder (see Fig.\ref{fig:spec} and Fig.\ref{fig:LC}). In the clumpy CSM scenario (Fig.~\ref{fig:schematic_picture}), the first rebrightening is caused by interaction with a major clump near the observer, producing the $\sim -4000$ km s$^{-1}$ excess in H$\alpha$, while the second arises from a far-side clump, causing the $\sim +2000$ km s$^{-1}$ excess. These clumps are randomly distributed, not bipolar, as shown by the enhancement of the different velocity components of the H$\alpha$ profile. Smaller clumps or a more uniform CSM explain the steady parts of the H$\alpha$ profile during both rebrightenings. The reddening and increased H$\alpha$/H$\beta$ absorption at the start of the second rebrightening indicate greater extinction and thus a shift of emitting regions from the near to far side of the ejecta.

Other observations also support the clumpy CSM scenario. The mismatch between spectral line velocities and SN brightness rules out a spherical photosphere expected from a spherical CSM. The photospheric spectrum indicate a photosphere with a temperature of $\sim 6000$ K even at phase +267 days. The H$\alpha$ line width, corresponding to a maximum velocity of $\sim 5000$ km s$^{-1}$, implies the emitting region has expanded to $\sim 10^{16}$ cm by that time. If this photosphere were spherical, its luminosity - assuming black-body radiation - would be $\sim 8 \times 10^{43}$ erg s$^{-1}$, or $\sim -21$ mag in optical bands, which is much brighter than the actual SN magnitude of $\sim -15$ mag at the same phase. Thus, the photosphere must be patchy. This is further supported by the $\sim 1$ \% polarization degree and the change in polarization angle (see Appendix~\ref{sec:appD}). The $\sim 90$-degree shift in polarization angle suggests that the CSM clumps lie in nearly perpendicular directions on the plane of the sky, relative to the progenitor.


The black lines in Fig.~\ref{fig:LC} show the $V$-band LCs from the model for the interaction between the SN ejecta and a spherical CSM presented in \citet[][]{Moriya2013}. Roughly speaking, the necessary mass-loss rate for explaining the brightness of SN~2024qiw is on the order of $\dot{M} \sim 10^{-2}$ M$_{\odot}$ yr$^{-1}$. In addition, since the durations of the first and second rebrightenings are $\sim50$ days, we can estimate the masses and sizes of the CSM clumps to be $\sim 10^{-1}$ M$_{\odot}$ and $\lesssim 2 \times 10^{15}$ cm, respectively, with the assumption of the shock velocity of 5000 km s$^{-1}$. Since the durations can be artificially extended by the diffusion effects in the shock regions, the actual clump masses and sizes can be slightly smaller than these values. Based on the timing of the start of the rebrightenings at phases +35 and +125 days and the durations of $\sim50$ days, we can infer the timings of these major mass ejections in the evolution history of the progenitor as $\sim 5-12$ and $\sim 17-24$ years before the explosion, with the assumption of $v_{\rm{shock}}=5000$km s$^{-1}$ and $v_{\rm{csm}}=100$ km s$^{-1}$. Here, the assumed CSM velocity falls within the range observed in Luminous Blue Variables (LBVs) \citep[$\sim100-1000$ km s$^{-1}$, e.g.,][]{Smith2017b}. While the estimated masses and sizes of the CSM clumps are not affected by the assumed CSM velocity, the inferred timing of the mass ejections is inversely proportional to the assumed CSM velocity.

These findings present a significant challenge to the standard stellar evolution theory, which do not predict such giant outbursts of clumpy material in the final decades before the core collapse. The inferred mass-loss rates, spatial clumpiness, and timing of the ejections are inconsistent with steady winds or binary-driven mass loss. One possible interpretation is that a an LBV star exploded while in a giant eruption phase — a scenario not permitted in classical models, which assume that LBVs evolve into Wolf–Rayet stars before the explosion \citep[e.g.,][]{Smith2017b}. LBVs are known to undergo episodic outbursts with durations ranging from days to several decades \citep[][]{Smith2011}, and can exhibit extreme mass-loss rates of $\sim 10^{-2}$ to $\sim 1$ M$_{\odot}$ yr$^{-1}$ during these phases \citep[][]{Smith2014}, consistent with the properties inferred for the CSM of SN~2024qiw. Alternatively, our results may indicate that another class of hydrogen-rich massive stars, distinct from classical LBVs, can also undergo similar energetic outbursts of clumpy material in the terminal stages of evolution. In either case, current stellar evolution theory must be revised to account for such mysterious, violent, and asymmetric mass loss shortly before explosion in hydrogen-rich massive stars.

Our results further challenge the widely claimed connection between LBVs and Type IIn SNe. In the case of SN~2024qiw, the explosion of an LBV-like progenitor did not result in a narrow-line-dominated Type II SN (i.e., Type IIn), but instead a broad-line-dominated Type II SN at least after $\sim 30$ days post-explosion. This behavior can be explained by the rapid enshrouding of the aspherical emitting regions by the SN ejecta. By contrast, many, though not all, Type IIn SNe continue to exhibit significant narrow Balmer lines at later epochs \citep[e.g.,][for reviews]{Smith2017a, Fraser2020, Dessart2024}. For the unshocked CSM regions responsible for these narrow lines to remain visible and not be obscured by the ejecta over a long period, the CSM must extend in the radial direction — e.g., as an extended envelope or in a disk-like or jet-like configuration. Indeed, in long-lived Type IIn SNe with disk-like CSM, narrow Balmer lines are observed for the first $\sim100$ days \citep[][]{Nagao2025}. Therefore, the discovery of SN~2024qiw suggests that explosions of LBV-like stars may be less common than previously thought. Another SN that may represent an explosion of an LBV-like star is ASASSN-13dn \citep[][]{Hueichapan2025}, a Type II SN that shows notable similarities to SN~2024qiw in both photometric and spectroscopic behavior.

Moreover, this conclusion raises questions about past LBV explosion candidates. Many previous studies have suggested LBV explosions as progenitors of Type IIn SNe, primarily based on the extreme inferred mass-loss rates \citep[e.g.,][]{Smith2017a}, detections of luminous progenitors, or evidence of pre-explosion variability \citep[e.g.,][]{Kotak2006, Gal-Yam2007, Smith2011, Smith2010, Pessi2022}. However, estimating SN progenitor masses from pre-explosion imaging is challenging. Outbursts can temporarily inflate luminosity, leading to mass overestimates, and the detected source may be a binary or stellar cluster. For example, the luminous pre-SN source of SN~2010jl was later identified as a young stellar cluster \citep[][]{Fox2017}. Similarly, not all pre-SN variability is necessarily linked to LBV-like activity. For example, SN2009ip exhibited luminous outbursts and a blue progenitor \citep[][]{Smith2010}, but its disk-like CSM suggests binary-driven mass ejection, rather than LBV eruptions \citep[][]{Mauerhan2013}. 

As the mass-loss mechanisms in LBVs are still under debate \citep[e.g.,][]{Smith2014}, the physical origin of the mass ejections seen in SN 2024qiw cannot be identified with certainty. Any viable scenario, however, must explain a hydrogen-rich massive star that exploded as a Type II SN and experienced clumpy, intensive mass loss at a rate of $\sim 10^{-2}$ M$_{\odot}$ yr$^{-1}$ during the final decades before core collapse. The distinct CSM geometries inferred for SN 2024qiw and for some Type IIn SNe further suggest that multiple, still-unknown mass-loss mechanisms operate in the late evolution of massive stars. 
The rarity of SN 2024qiw could reflect an intrinsically rare progenitor, or alternatively, a more common progenitor that only rarely undergoes massive pre-explosion ejections. In the former case, such progenitors would need relatively uniform properties to produce LBV-like mass loss in the final decades, leading to similar explosion characteristics. However, the somewhat different photospheric and spectroscopic properties of ASASSN-13dn \citep[][]{Hueichapan2025} compared with SN 2024qiw might argue against this scenario. If instead these intense mass ejections occur at varying times, similar progenitors could produce different SN types. Given that the early photometric and spectroscopic properties of SN 2024qiw resemble those of typical Type IIL SNe, it might have appeared as a Type IIL event had its massive CSM clumps been ejected earlier. Conversely, if mass loss had persisted longer before explosion, the event might have appeared as a stripped-envelope SN (SESN), or a SESN interacting with dense clumpy CSM. SESNe with bumpy light curves \citep[e.g.,][]{Sollerman2020} may thus represent explosions of progenitors similar to that of SN 2024qiw but occurring at different evolutionary phases.


\begin{acknowledgements}

We thank Philip D. Michel for kindly providing a spectrum of SN~2023ixf. This work is partly based on observations made under program IDs P70-016 and P70-018 with the Nordic Optical Telescope, owned in collaboration by the University of Turku and Aarhus University, and operated jointly by Aarhus University, the University of Turku and the University of Oslo, representing Denmark, Finland and Norway, the University of Iceland and Stockholm University at the Observatorio del Roque de los Muchachos, La Palma, Spain, of the Instituto de Astrofisica de Canarias. The Gravitational-wave Optical Transient Observer (GOTO) project acknowledges support from the Science and Technology Facilities Council (STFC, grant numbers ST/T007184/1, ST/T003103/1, ST/T000406/1, ST/X001121/1 and ST/Z000165/1) and the GOTO consortium institutions; University of Warwick; Monash University; University of Sheffield; University of Leicester; Armagh Observatory \& Planetarium; the National Astronomical Research Institute of Thailand (NARIT); University of Manchester; Instituto de Astrofísica de Canarias (IAC); University of Portsmouth; University of Turku. T.N. and H.K. acknowledge support from the Research Council of Finland projects 324504, 328898, 340613, and 353019. K.M. acknowledges support from JSPS KAKENHI grant (JP24KK0070, JP24H01810) and the JSPS Open Partnership Bilateral Joint Research Projects between Japan and Finland (JPJSBP120229923). S.M. acknowledges support from the Research Council of Finland project 350458. T.L.K. acknowledges a Warwick Astrophysics prize post-doctoral fellowship, made possible thanks to a generous philanthropic donation. D.S. acknowledges support from the UK Science and Technology Facilities Council through grant numbers ST/T007184/1, ST/T003103/1, and ST/T000406/1.

\end{acknowledgements}

  \bibliographystyle{aa} 
  \bibliography{aa.bib} 

\appendix

\section{Discovery, explosion epoch, redshift, and extinction} \label{sec:appA}

\begin{figure}
   \centering
            \includegraphics[width=\hsize]{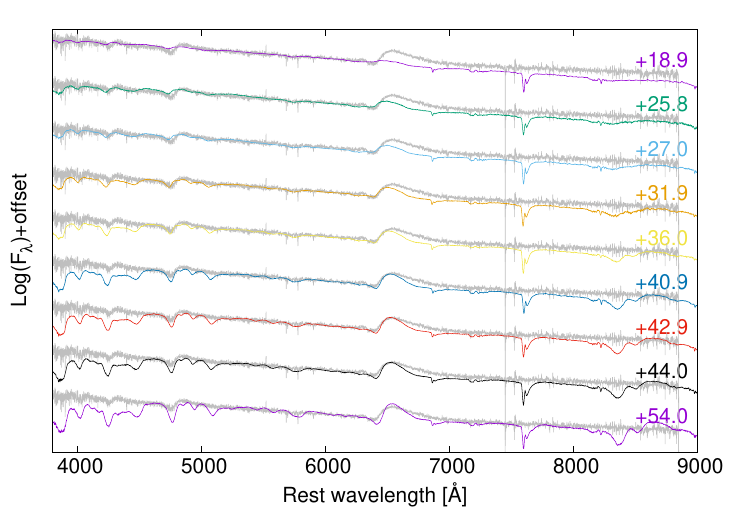}
      \caption{Spectral comparison between SNe~2024qiw and 2023ixf. The early spectra of SN~2023ixf (colored lines) are overplotted on the earliest (Phase~+27) spectrum of SN~2024qiw (gray). The phases are given relative to the explosion.
              }
      \label{fig:app1}
   \end{figure}

SN~2024qiw was discovered by the Asteroid Terrestrial-impact Last Alert System \citep[ATLAS;][]{Tonry2018,Smith2020}, which was reported with the discovery date of 28.41 July 2024 UT \citep[60519.41 MJD;][]{Tonry2024}. According to the ATLAS forced photometry server \citep[][]{Shingles2021}, the actual first detection was on 60513.18 MJD. It was classified as a Type~II SN two days after the discovery report \citep[][]{Martin2024}. The last non-detection was on 12.99 April 2024 (60412.99 MJD) before the solar conjunction, which is $\sim 100$ days before the first detection \citep[][]{Tonry2024}. We estimate the explosion date of SN~2024qiw by comparing its earliest spectrum with those of the Type~IIL SN~2023ixf, which is photometrically and spectroscopically similar to SN~2024qiw at early phases before the first rebrightening (see Section~\ref{sec:result}). Fig.~\ref{fig:app1} shows spectral comparison between SNe~2024qiw and 2023ixf. Based on the color and the flux ratios between the continuum and Balmer lines, we judge that the earliest epoch spectrum of SN~2024qiw is most similar to that of SN~2023ixf at $+27.0$ days after the explosion. Thus, we assume the explosion date of SN~2024qiw to be 60495 MJD. Throughout the paper, we measure the epochs from this date, unless otherwise mentioned. 
The redshift of the host galaxy (WISEA J045257.31-053207.2) is $z =0.010280 \pm 0.00002$ \citep[][]{Theureau1998}, which is reported in the NASA/IPAC Extragalactic Database (NED)\footnote{\url{https://ned.ipac.caltech.edu/}}. The corresponding distance modulus under the assumption of $H_0$=73 km s$^{-1}$ Mpc$^{-1}$, $\Omega_M$ = 0.27 and $\Omega_{\Lambda}$ = 0.73 is $\mu=33.15$.
Since there is no significant Na~I~D interstellar absorption at the host-galaxy redshift in our spectra, we do not consider the extinction within the host galaxy and take only the MW extinction into account \citep[$E(B-V)=0.033$][]{Schlafly2011} assuming $R_V$ = 3.1 and the extinction curve by \citet[][]{Cardelli1989}.

\section{Observations and data reduction}  \label{sec:appB}

Here, we present the details on our photometric, spectroscopic, and polarimetric observations for SN~2024qiw. The logs of the observations are provided in Tables~\ref{tab:spec_24qiw} and \ref{tab:pol_24qiw}.

\subsection{Photometry}

We retrieved $L$-band light curves from the Gravitational-wave Optical Transient Observer \citep[GOTO;][]{Steeghs2022, Dyer2024} via the GOTO forced photometry service (Jarvis et al in prep).  The GOTO $L$-band is a wide filter, approximately equivalent to $g$+$r$.
We also collected the orange- and cyan-band light curves obtained by ATLAS through the ATLAS Forced Photometry server \citep[][]{Shingles2021}.

\subsection{Spectroscopy}

We obtained optical spectra of SN~2024qiw using the Alhambra Faint Object Spectrograph and Camera (ALFOSC)\footnote{\url{https://www.not.iac.es/instruments/alfosc/}} mounted on the 2.56-m Nordic Optical Telescope (NOT)\footnote{\url{https://www.not.iac.es/}} at the Roque de los Muchachos Observatory. We adopted Grism 4, which has a wavelength coverage of 3200–9600 {\AA} and a spectral resolution of $\sim 360$. We reduced the spectra using the alfoscgui pipeline\footnote{FOSCGUI is a graphical user interface aimed at extracting SN spectroscopy and photometry obtained with FOSC-like instruments. It was developed by E. Cappellaro. A package description can be found at \url{https://sngroup.oapd.inaf.it/foscgui.html}}, which includes the overscan, bias, and flat-field corrections, the cosmic-ray removal, the extraction of of a one-dimensional spectrum, and the sky subtraction. We calibrate the wavelength scale of the extracted spectra using an arc lamp, and the flux scale using a sensitivity function derived from a standard star observed on the same night. We used the classification spectrum reported in the Transient Name Server \citep[][]{Martin2024}.

\subsection{Polarimetry}

We conducted $V$-, $R$- and $i$-band imaging polarimetry of SN~2024qiw using ALFOSC/NOT. For the linear polarimetry, we used a half-wave plate (HWP) with 4 HWP angles ($0^{\circ}$, $22.5^{\circ}$, $45^{\circ}$ and $67.5^{\circ}$) and a calcite plate. We reduced and analyzed the data adopting the standard methods, e.g., in \citet[][]{Patat2017}, and using IRAF\citep[][]{Tody1986,Tody1993}. After correcting the bias and flat-field in all the frames, we performed aperture photometry on the ordinary and the extraordinary components of the transient for the four HWP angles. Here, we adopted an aperture size that is twice as large as the full width at half maximum (FWHM) of the ordinary beam's point-spread function, and the sky region from the outer edge of the aperture to the radius that is twice as large as the aperture. From these measurements, we calculated the Stokes parameters, the polarization degrees and the polarization angles. In the calculations of the polarization degrees, we subtracted the polarization bias following the treatment in \citet[][]{Wang1997}.

\begin{table}
  \caption{Log of the spectroscopic observations of SN~2024qiw.}
  \label{tab:spec_24qiw}
  \centering
  \begin{tabular}{lcccc}
    \hline\hline
    Date (UT) & MJD (days) & Phase (days) & Exp. time (s) \\
    \hline
    2024-10-02.18 & 60585.18 & +90  &  600  \\
    2024-10-12.15 & 60595.15 & +100 &  900  \\
    2024-10-31.18 & 60614.18 & +119 &  900  \\
    2024-11-14.02 & 60628.02 & +133 &  900  \\
    2024-11-25.16 & 60639.16 & +144 &  900  \\
    2024-12-28.95 & 60672.95 & +178 &  900  \\
    2025-01-30.93 & 60705.93 & +211 &  1200 \\
    2025-03-27.85 & 60761.85 & +267 &  1800 \\
    \hline
  \end{tabular}
  
  \begin{minipage}{.88\hsize}
    \smallskip
    \emph{Notes:} The phase is measured relative to the assumed explosion date (60495 MJD).
  \end{minipage}
\end{table}

\begin{table*}
  \caption{Log and measurements of the polarimetric observations of SN~2024qiw.}
  \label{tab:pol_24qiw}
  \centering
  \begin{tabular}{lcccccc}
    \hline\hline
    Date (UT) & MJD (days) & Phase (days) & Exp. time (s) & Pol. degree (\%) & Pol. angle (deg) & Filter \\
    \hline
    \multirow{2}{*}{2024-10-02.18} & \multirow{2}{*}{60585.18} & \multirow{2}{*}{+90}  & $4 \times 80$  & $1.56 \pm 0.41$ & $0.7 \pm 7.1$   & R \\
                                   &                          &                    & $4 \times 80$  & $(0.00 \pm 0.63)$ & $(168.5 \pm 33.4)$ & i \\
    \hline
    \multirow{3}{*}{2024-10-12.13} & \multirow{3}{*}{60595.13} & \multirow{3}{*}{+100} & $4 \times 100$ & $1.11 \pm 0.34$ & $63.7 \pm 8.0$   & V \\
                                   &                          &                    & $4 \times 100$ & $(0.00 \pm 0.28)$ & $(23.0 \pm 31.6)$  & R \\
                                   &                          &                    & $4 \times 100$ & $0.52 \pm 0.43$ & $155.5 \pm 16.0$ & i \\
    \hline
    \multirow{2}{*}{2024-10-31.19} & \multirow{2}{*}{60614.19} & \multirow{2}{*}{+119} & $4 \times 100$ & $1.04 \pm 0.39$ & $90.9 \pm 9.5$   & R \\
                                   &                          &                    & $4 \times 100$ & $(0.13 \pm 0.57)$ & $(91.9 \pm 25.6)$ & i \\
    \hline
    \multirow{3}{*}{2024-11-25.17} & \multirow{3}{*}{60639.17} & \multirow{3}{*}{+144} & $4 \times 100$ & $0.57 \pm 0.50$ & $95.7 \pm 16.7$  & V \\
                                   &                          &                    & $4 \times 100$ & $1.03 \pm 0.42$ & $116.8 \pm 10.3$ & R \\
                                   &                          &                    & $4 \times 100$ & $2.08 \pm 0.67$ & $78.3 \pm 8.5$   & i \\
    \hline
    2025-01-02.98                  & 60677.98                 & +183                & $4 \times 100$ & $(0.00 \pm 0.48)$ & $(31.2 \pm 40.0)$ & R \\
    \hline
  \end{tabular}

  \begin{minipage}{.88\hsize}
    \smallskip
    \emph{Notes:} The phase is measured relative to the assumed explosion date (60495 MJD).
  \end{minipage}
\end{table*}

\section{Spectral comparison between SN~2024qiw and other Type~II SNe} \label{sec:appC}

\begin{figure}
   \centering
            \includegraphics[width=\hsize]{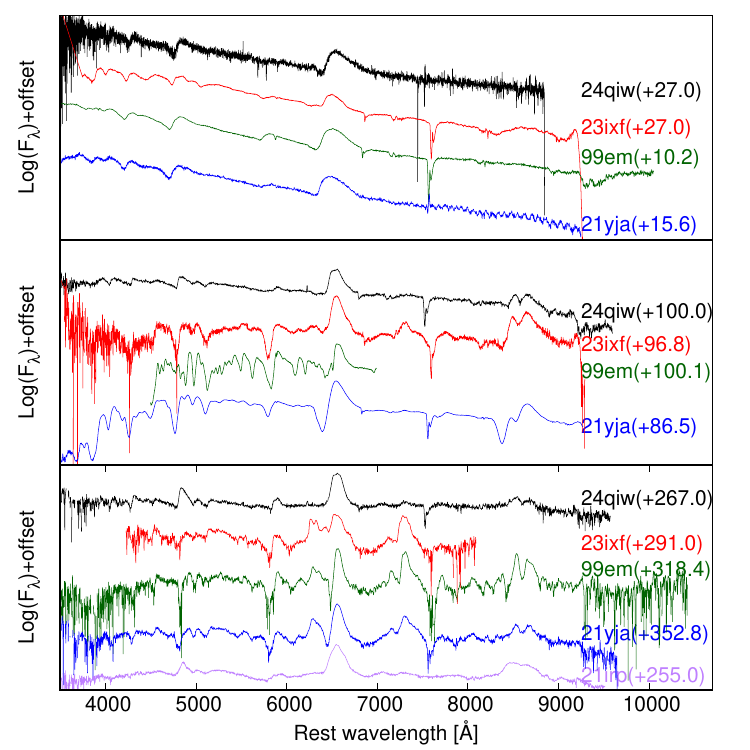}
      \caption{Spectral comparison with other Type~II SNe at three different phases. The data for the comparison SNe were taken from \citet[][]{Singh2024, Michel2025} for SN~2023ixf, \citet[][]{Hamuy2001, Leonard2002} for SN~1999em, \citet[][]{Hosseinzadeh2022, Nagao2024} for SN~2021yja, and \citet[][]{Reynolds2025a} for SN~2021irp, via the Weizmann Interactive Supernova Data Repository \citep[WISeREP;][]{Yaron2012}.
      }
      \label{fig:spec_comp}
   \end{figure}

Fig.~\ref{fig:spec_comp} shows spectral comparison with other Type II SNe.
At early phases, the spectrum of SN~2024qiw is similar to those of other Type IIP/IIL SNe at the peak brightness. It is noted that this similarity between SN~2024qiw and Type IIP and IIL SNe supports the validity of our estimation for the explosion epoch for SN~2024qiw. At $\sim 100$ days after the explosion, which roughly corresponds to the end of the plateau phase for Type IIP/IIL SNe, the spectral difference between SN~2024qiw and the Type IIP SNe becomes clear. SN~2024qiw shows minimal absorption parts for the lines, while the Type IIP SNe have strong absorption parts (e.g., in Balmer lines, Ca II triplet, Na I D line, Fe II lines). On the other hand, the spectrum of the Type IIL SN~2023ixf is relatively alike to that of SN~2024qiw, although the strong absorption of the Na I D line does not match. At several hundreds of days after the explosion, when Type IIP/IIL SNe show the so-called nebular spectrum, the spectrum of SN~2024qiw does not anymore resemble that of the Type IIL SN~2023ixf. It more closely resembles the spectrum of SN~2021irp, which shows broad permitted lines with emission-dominated profiles superimposed on a continuum, along with the so-called Fe bump at blue wavelengths ($\lambda \lesssim 5500$ {\AA}).

\section{Polarimetric properties of SN~2024qiw} \label{sec:appD}


\begin{figure}
   \centering
            \includegraphics[width=\hsize]{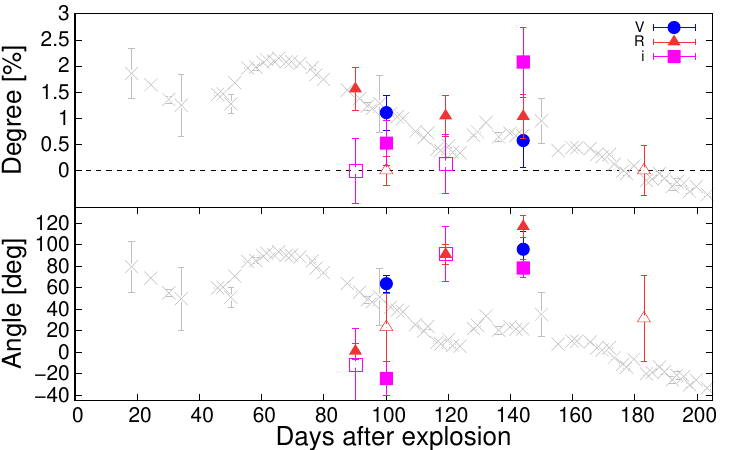}
      \caption{The evolution of the degree and angle of the multi-band polarization of SN 2024qiw. The open symbols show the data that are consistent with zero polarization, while the filled symbols show the data that exhibit non-negligible polarization. The gray points show its $o$-band light curve.
              }
      \label{fig:pol}
   \end{figure}

Fig.~\ref{fig:pol} shows the time evolution of the multi-band polarization of SN~2024qiw. Although the data are noisy and several measurements are consistent with zero polarization within the uncertainties, the polarization degree is $\sim 1$ \%, with time-variable polarization angles ranging from $\sim 0$ (during the first rebrightening) to $\sim 90$ degrees (around the peak of the second rebrightening). The empirical relation between dust extinction and resulting polarization \citep[$P_{\rm{ISP}} \lesssim 9E(B-V)$;][]{Serkowski1975} suggests that the interstellar polarization (ISP) should be less than $\sim 0.3$ \% for the assumed extinction of $E(B-V)=0.033$ mag. Therefore, the observed polarization likely reflects aspherical distributions in the continuum-emitting and/or line-absorbing regions, as well as their temporal evolution.

\end{document}